\documentclass[fleqn,twoside]{article}
\usepackage{espcrc2}
\usepackage{cite}
\usepackage{axodraw}
\usepackage{amsmath}
\usepackage{eufrak}

\newcommand{\bq}{\\begin{equation}}  
\newcommand{\eq}{\end{equation}}
\newcommand\beq{\begin{equation}}   
\newcommand\eeq{\end{equation}}
\newcommand\bea{\begin{eqnarray}}
\newcommand\eea{\end{eqnarray}}

\usepackage{graphicx}
\usepackage[figuresright]{rotating}

\title{
Computation of Gr\"obner Bases for Two-Loop Propagator Type
Integrals}
\author{O.V. Tarasov\address{DESY Zeuthen, 
Platanenallee 6, D-15738 Zeuthen, Germany}
\thanks{On leave of absence from JINR,
141980 Dubna (Moscow Region), Russian Federation.} }
\begin{document}
\topmargin -1cm

\begin{abstract}
\noindent
The Gr\"obner basis technique for calculating Feynman diagrams
proposed in \cite{ovt98} is applied to the two-loop propagator 
type integrals with arbitrary masses and momentum. 
We describe the derivation of Gr\"obner bases for all integrals with  
1PI topologies and  present  elements of  the Gr\"obner bases.
\vspace{1pc}
\end{abstract}

\maketitle

\section{INTRODUCTION}
\noindent
The most efficient algorithms to evaluate Feynman diagrams 
exploit recursive methods based on integration by parts technique 
\cite{ibpm} or  technique of generalized recurrence relations 
\cite{ovt96},\cite{ovt97}.  It is easy to derive hundreds of recurrence 
relations but it is by far  not so easy to formulate optimal algorithm 
how to use them for the reduction of a given type of integrals to a minimal
set of basis integrals. 
In \cite{ovt98} it was proposed to use  the Gr\"obner basis  method 
\cite{Buchberger} as a mathematical  background to solve this problem.  
This proposal can be realized in different ways. For example,
the system of recurrence relations can be rewritten as  a system 
 of partial differential equations (PDE). The 
 set of  relations needed for reduction of Feynman integrals with 
 different powers of propagators to the set of basis integrals 
 will be Gr\"obner basis of this overdetermined system 
 of differential equations.
Information about the minimal basis of integrals also 
will be contained in the Gr\"obner basis.
Another possibility will be to rewrite recurrence relations
in terms of operators shifting powers of propagators.
Each equation should be  considered as an operator.
Operators shifting powers of propagators satisfy an Ore algebra' 
condition and therefore,
to find minimal set of recurrence relations from our system of
equations ammounts to computation of it's Gr\"obner basis 
in Ore algebra. 
 
 Our preliminary investigation reveals that both approaches
 have some merits and shortcomings. The first approach 
 certainly will be preferable in case when integrals
 depend on many mass scales. 
 For integrals depending on two - three mass scales the second 
 approach can be more efficient.
 
 In this article we present the computation of the Gr\"obner basis 
 for the two-loop propagator type  integrals with arbitrary
 momentum and masses. The derivation is based on a representation
 of recurrence relations as partial differential equations.
 For integrals with three, four and five lines we used notation
 $J_3^{d}$, $V^{d}$  and $F^{d}$ respectively which was
 adopted in our paper \cite{ovt97}.

\section{TWO-LOOP `SUNSET' INTEGRALS}

We begin our consideration with the so-called, `sunset' type
of integrals given on Fig.1.
\noindent
\vspace{-5mm}
\begin{center}
\begin{picture}(140,90)(0,0)
\CArc(70,50)(25,0,180)
\CArc(70,50)(25,180,360)
\Line(30,50)(110,50)
\Vertex(45,50){2}
\Vertex(95,50){2}
\Text(45,72)[]{$1$}
\Text(70,58)[]{$2$}
\Text(70,35)[]{$3$}
\Text(70,10)[]{${\rm ~Fig.1}$}
\end{picture}
\end{center}
\vspace{-3mm}
Recurrence relations for this type of integrals can be obtained
from two equations \cite{tHV},\cite{ibpm}
\begin{eqnarray}
\label{K1}
\int d^d k_1 d^dk_2~~ \frac{\partial }{\partial { k_{1\mu}}}
~~\frac{S_{\mu \nu \rho} Q_{\rho \nu} }
{c_1^{\nu_1}c_2^{\nu_2}c_3^{\nu_3}} \equiv 0,
\\
\int d^d k_1 d^dk_2~~ \frac{\partial }{\partial { k_{2\mu}}}
~~\frac{ S_{\mu \nu \rho} Q_{\rho \nu}}
{c_1^{\nu_1}c_2^{\nu_2}c_3^{\nu_3}} \equiv 0,
\label{K2}
\end{eqnarray}
where
\begin{eqnarray*}
&&c_1=k_1^2-m_1^2,~~c_2=(k_1-k_2)^2-m_2^2, \\
&&c_3=(k_2-q)^2-m_3^2, \\
&&S_{\mu \nu \rho} = w_1g_{\mu \nu} q_{\rho}+  \\
&& ~~~~w_2g_{\rho \nu}q_{\mu} 
 + w_3g_{\mu \rho} q_{\nu} + w_4q_{\mu}q_{\nu}q_{\rho}, 
\end{eqnarray*}
and $Q_{\mu \nu} $ is an arbitrary tensor. In our consideration
we restrict ourselves to the second rank tensor in $k_1,k_2$ and
external momentum $q$:
\begin{eqnarray}    
Q_{\mu \nu} = x_1 { k_{1 \nu} k_{1 \mu}} 
+ x_2 {k_{1 \nu} k_{2 \mu}} 
+ x_3 {k_{1 \mu} k_{2 \nu}}
\nonumber \\
+ x_4 {k_{2 \nu} k_{2 \mu}}
+ x_5 {k_{1 \nu}} q_{ \mu} 
+ x_6 {k_{1 \mu}} q_{ \nu}
\\
+ x_7 {k_{2 \nu}} q_{ \mu} 
+ x_8 {k_{2 \mu}} q_{ \nu} 
+ x_9 q_{ \mu} q_{ \nu}.\nonumber
\end{eqnarray}
Parameters $x_i$,$w_j$ are arbitrary constants.
After differentiation integrals with irreducible numerators
$$
\int \int d^dk_1 d^dk_2 \frac{{  (k_2q)^{\alpha} (k_1k_2)^{\beta}}}
{c_1^{\nu_1}c_2^{\nu_2}c_3^{\nu_3}}
$$
were expressed in terms of integrals with shifted dimension.
For example,
\begin{eqnarray}
&&\int \!\! \int \!\!d^dk_1 d^dk_2
 \frac{(k_1q)(k_2q)}{c_1c_2c_3}\!=\nonumber \\
&& 
\frac{q^2}{2\pi^2} \int \!\! \int \!\!
\frac{d^{d+2}k_1 d^{d+2}k_2}{c_1c_2^2c_3}
 \\
&&
+\frac{q^4}{\pi^4}
\int \!\! \int \!\! \frac{d^{d+4}k_1 d^{d+4}k_2}
{c_1  c_2^2 c_3^3}
\left[
\frac{2}{c_1 }
+\frac{4}{c_2}            \right]. \nonumber
\label{examp}
\end{eqnarray}
Propagators with 'shifted' indices should be represented as:
\begin{equation}
\frac{1}{c_j^{1+r}}=\frac{1}{r!}
\frac{\partial^r}{\partial z_j^{r}}
\frac{1}{c_j},
\end{equation}
where
$$
z_j = m_j^2.
$$
`Sunset' type integrals $J_3^{d}$ with different shifts of the parameter 
of the space-time dimension $d$ were considered as different functions, i.e:
\begin{equation}
J_3^{d}={\rm sun0}, ~~~
J_3^{d+2}={\rm sun1}, ~~~
J_3^{d+4}={\rm sun2},
 \ldots
\end{equation}
These substitutions allows one to transform
the system of recurrence relations into a linear system of
PDEs for the vector function $P \equiv \{sun0, sun1, \ldots \}$.
 At $\nu_1=\nu_2=\nu_3=1$
Eqs. (\ref{K1}),(\ref{K2}) give 28 different relations connecting
`sunset' integrals with different shifts of the parameter of space-time
dimension $d$ and different products of one-loop tadpole type integrals.

To find the Gr\"obner basis of our system of PDE  we used package
${\rm Rif}$ \cite{Rif} which is distributed as part of Maple.

Just for illustration we give a concrete example of the program
written in Maple for computation of the Gr\"obner basis.

\begin{verbatim}
read `equations`:
eq29:=z1*diff(T1(z1),z1) 
                 - (d/2-1)*T1(z1)=0:
eq30:=z2*diff(T2(z2),z2) 
                 - (d/2-1)*T2(z2)=0:
eq31:=z3*diff(T3(z3),z3) 
                 - (d/2-1)*T3(z3)=0:
syst:={seq(eq||j,j=1..31)}:
syst:= syst union {eq29,eq30,eq31}:
syst:= convert(syst,list):

with(DEtools):

invr := [z1,z2,z3]:
dvars:=[sun3,sun2,sun1,sun0,T1,T2,T3]:
Rnk:=[[0,0,0,150,100,50,0,0,0,0],
      [1,1,1,  0,  0, 0,0,0,0,0]]:
bas := rifsimp(
  syst,dvars,indep=invr,ranking=Rnk);
GBasis:= bas[Solved];
quit;
\end{verbatim}
First, the system of 28 differential equations stored in the 
file `equations' is read in  and then we add to it three 
differential relations for one-loop tadpole integrals
with different masses: 
$$
\frac{\partial}{\partial z_j} T_j(z_j)=\frac{d-2}{2z_j} T_j(z_j),~~~~
T_j(z_j) = \int \frac{d^d k_1}{k_1^2-z_j}.
$$
In order to get rid of integrals with shifts of $d$
we adopted here a ranking giving  weight proportional
to the shift of space time dimension. Also derivatives
of functions have weight higher than the function itself.

It took 5 min on  a 1.6 GHz PC to obtain the Gr\"obner basis
consisting of 19 differential relations for $J_3^d$ with
different shifts of $d$ and three relations for $T_i(z_i)$.
The left hand sides of these 19 relations are:
\begin{eqnarray}
&&
\frac {\partial ^{2}
{\it J_3^{{  d}}}}
{\partial {{\it z_1}}^{2}},~~
{\frac {\partial ^{2}
{\it J_3^{{  d}}} 
}{\partial {\it z_1}\partial {\it z_2}}},~~
{\frac {\partial ^{2}
{\it J_3^{{  d}}} 
}{\partial {\it z_1}\partial {\it z_3}}},~~
{\frac {\partial ^{2}
{\it J_3^{{  d}}} 
}{\partial {{\it z_2}}^{2}}},~~
\nonumber \\
&&
{\frac {\partial ^{2}
{\it J_3^{{  d}}} 
}{\partial {\it z_2}\partial {\it z_3}}}
,~~ 
{\frac {\partial ^{2}
{\it J_3^{{  d}}} 
}{\partial {{\it z_3}}^{2}}}
,
\nonumber \\
&& \nonumber \\
&& \nonumber \\
&&
{\frac {\partial ^{2}
{\it J_3^{{  d+2}}} 
}{\partial {\it z_1}\partial {\it z_2}}},~~ 
{\frac {\partial ^{2}
{\it J_3^{{  d+2}}} 
}{\partial {\it z_1}\partial {\it z_3}}}
,~~~
{\frac {\partial ^{2}
{\it J_3^{{  d+2}}} 
}{\partial {\it z_2}\partial {\it z_3}}}
,
\nonumber 
\\
&& \nonumber \\
&& \nonumber \\
&&
{\frac {\partial ^{5}
{\it J_3^{{  d+4}}} 
}{ \partial {{\it z_3}}^2 \partial {{\it z_1}}^{3}  }}
,~~
{\frac {\partial ^{5}
{\it J_3^{{  d+4}}} 
}{ \partial {{\it z_1}}^{2}  \partial {{\it z_3}}^{3} }  },~~
{\frac {\partial ^{5}
{\it J_3^{{  d+4}}} 
}{\partial {\it z_1} \partial {\it z_2} \partial {{\it z_3}}^{3}}},
 \nonumber \\
&& \nonumber \\
&&
{\frac {\partial ^{5}
{\it J_3^{{  d+4}}} 
}{\partial {\it z_3}\partial {{\it z_2}}^{3}
\partial {\it z_3}}},~~
{\frac {\partial ^{5}
{\it J_3^{{  d+4}}} 
}{\partial {{\it z_3}}^{2}\partial {{\it z_2}}^{2}
\partial {\it z_3}}},~~
{\frac {\partial ^{4}
{\it J_3^{{  d+4}}} 
}{\partial {{\it z_1}}^2  \partial {\it z_2}
 \partial {\it z_3} }}, \nonumber \\
&&
{\frac {\partial ^{4}
{\it J_3^{{  d+4}}} 
}{\partial {\it z_1}  \partial {{\it z_2}}^2
  \partial {\it z_3} }},
{\frac {\partial ^{7}
{\it J_3^{{  d+6}}} 
}{\partial {{\it z_1}}^2 \partial {{\it z_2}}^2 \partial {{\it z_3}}^3   }}
,~~
\nonumber \\
&&
{\frac {\partial ^{7}
{\it J_3^{{  d+6}}} 
}{\partial {{\it z_1}}^3 \partial {{\it z_2}} \partial {{\it z_3}}^3   }}
,~~
{\frac {\partial ^{7}
{\it J_3^{{  d+6}}} 
}{\partial {{\it z_1}} \partial {{\it z_2}}^3 \partial {{\it z_3}}^3   }}.
\label{basis28}
\end{eqnarray}
In complete agreement with \cite{ovt97} the number of equations 
for $J_3^d$ without shifts of $d$
is enough to reduce integrals with $\nu_1+\nu_2+\nu_3>4$ to
four basis integrals $J_3^d, \partial J_3^d/ \partial z_1, 
\partial J_3^d/ \partial z_2, \partial J_3^d/ \partial z_3$,
and three products of one-loop tadpoles.
The basis for integrals with shifted $d$ is not complete.
As was shown in \cite{ovt97} these integrals can be expressed
in terms of $J_3^{d}$ and its first derivatives.
Such a relation was not obtained from the 28 equations produced by
Eqs. (\ref{K1})-(\ref{K2}). It turns out that if we shift
$d \rightarrow d+2$ in (\ref{K1}),(\ref{K2}) and add the
obtained 28 equations  to original 28 equations then from this 
enlarged system we will obtain in the Gr\"obner basis the required 
expression for $J_3^{d+2}$ which agrees  with \cite{ovt97}.
Therefore, computing Gr\"obner basis we must include these equations
in our system from  the very beginning.

Taking into account the relation for $J_3^{d+2}$ in terms
of  $J_3^d, \partial J_3^d/ \partial z_1, 
\partial J_3^d/ \partial z_2, \partial J_3^d/ \partial z_3$,
we will find that  the Gr\"obner basis for `sunset' 
integrals consists out of 10 relations: 6 relations with second 
derivatives for $J_3^d$,
one relation for $J_3^{d+2}$ and three relations for one loop
tadpole integrals. Expressions for derivatives of $J_3^{d+2}$,
$J_3^{d+4}$, $J_3^{d+6}$ will be excluded from (\ref{basis28}).

It should be noted that keeping in $Q$ only linear terms in $k_1,k_2$ 
gives incomplete basis for $J_3^d$.

By using the Gr\"obner basis reduction of $J_3^{d}$, $J_3^{d+2} {\ldots} $
integrals with $\nu_1+\nu_2+\nu_3>4$ can be done with elimination commands
built in Maple. We wrote our own procedure for such a reduction
taking into account specific properties of our Gr\"obner basis. 
This procedure works essentially faster than built in tools of Maple.

\section{INTEGRALS $V^{d}$ }
\noindent
In the same manner as it was done for the `sunset' integrals
we repeated computation of the Gr\"obner basis for 
propagator type integrals with the topology given in Fig.2.
\begin{center}
\begin{picture}(140,70)(0,0)
\CArc(70,40)(25,0,180)
\CArc(70,40)(25,180,360)
\CArc(45,15)(25,0,90)
\Line(30,40)(45,40)
\Line(95,40)(110,40)
\Vertex(95,40){2}
\Vertex(45,40){2}
\Vertex(70,15){2}
\Text(42,57)[]{$2$}
\Text(42,27)[]{$3$}
\Text(97,27)[]{$4$}
\Text(68,38)[]{$1$}
\Text(70,0)[]{${\rm Fig.~2.}$}
\end{picture}
\end{center}
The  Gr\"obner basis consists of 
ten relations which were already obtained for
`sunset' integrals,
five relations for $V^{{d}}$ and
three relations for 1-loop propagator
integrals 
\begin{equation} 
 G^{ d }=\int \frac{d^d k_1}{[k_1^2-z_2][(k_1-q)^2-z_4]}.
\end{equation} 
New  relations  needed in the Gr\"obner basis are:
\begin{eqnarray*}
&&{\frac {\partial
{\it V^{{  d}}}
 }{\partial {\it z_1}}}
,~~~
{\frac {\partial
{\it V^{{  d}}}
 }{\partial {\it z_2}}}
,~~~
{\frac {\partial
{\it V^{{  d}}}
 }{\partial {\it z_3}}}
,~~~
{\frac {\partial 
{\it V^{{  d}}
 }}{\partial {\it z_4}}}
,~~~
{\it V^{{  d+2}}}
 ,
\nonumber \\
&&
{\frac {\partial  {\it G^{ d}}}{\partial {\it z_2}}},~~~
{\frac {\partial  {\it G^{{d}}} }{\partial {\it z_4}}},
~~~~~G^{{  d+2}}.
\end{eqnarray*}
As it was for the `sunset' case the number of elements
in the Gr\"obner basis and the number of basic integrals
for $V$ type integrals is in agreement
with \cite{ovt97}.

\section{INTEGRALS   $F^{d}$ }
\noindent
Gr\"obner basis for $F^d$ integrals  with  the topology 
given in Fig.3 includes  basis of `sunset' type integrals,
$V$ type integrals and also new relations.
\begin{picture}(30,110)(90,50)
\CArc(190,110)(30,0,180)
\CArc(190,110)(30,180,360)
\Line(190,80)(190,140)
\ArrowLine(140,110)(160,110)
\ArrowLine(220,110)(240,110)
\Vertex(160,110){1.75}
\Vertex(190,80){1.75}
\Vertex(190,140){1.75}
\Vertex(220,110){1.75}
\Text(160,130)[]{1}
\Text(220,130)[]{2}
\Text(160,90)[]{3}
\Text(220,90)[]{4}
\Text(197,110)[]{5}
\Text(146,117)[]{$q$}
\end{picture}
\Text(70,15)[]{${\rm Fig.3.}$}
\vglue 1mm{
Six new relations for the following quantities 
\begin{equation}
{\frac {\partial{\it F^{{  d}}}
 }{\partial {\it z_1}}}
,~~
{\frac {\partial
{\it F^{{  d}}}}{\partial {\it z_2}}}
,~~
{\frac {\partial
{\it F^{{  d}}}
 }{\partial {\it z_3}}}
,~~
{\frac {\partial 
{\it F^{{  d}}}}{\partial {\it z_4}}}
,~~
{\frac {\partial 
{\it F^{{  d}}}}
{\partial {\it z_5}}}
,~~
{\it F^{{  d+2}}},
\end{equation}
appeared in the Gr\"obner basis of $F^{{d}}$  integrals.
Again these relations and the number of master integrals 
agree with the results given in \cite{ovt97}.

\section{CONCLUDING REMARKS}

The considered examples  demonstrate that with already existing 
software one can compute Gr\"obner bases for a rather
complicated type of Feynman integrals.
Our preliminary investigation reveals   also that  Gr\"obner
bases for vertex integrals with arbitrary masses and
external momenta can be computed with the existing
computers and available software. Corresponding 
expressions are rather long, sometimes thousands lines of codes.
However after the Gr\"obner basis is computed, it can be stored
and used in  different applications.
At the present time we cannot use Gr\"obner bases for
reduction of integrals with arbitrary kinematics.
The reason is that some masses or Gram determinants
appear in denominators and therefore one cannot make
reductions when these factors are zero.
Different solutions of this problem are possible
and they will be presented in future publications.

\noindent

\vspace*{1mm}
\noindent
{\bf Acknowledgment.}~~
I would like  to thank J. Fleischer and  V. Gerdt for useful discussions
and M. Kalmykov for useful remarks. I am also very thankful 
to Dr. G. Meyer zu Sieker for providing at my disposal a notebook 
which was used for all  calculations described in the paper.
This paper was supported  by DFG Sonderforschungsbereich Transregio~9, 
Computergest\"utzte Theoretische Physik.


\end{document}